\documentclass[aps,prd,twocolumn,showpacs,superscriptaddress, floatfix]{revtex4-2}  

\usepackage{graphicx}  
\usepackage{dcolumn}   
\usepackage{bm}        
\usepackage{amssymb}   
\usepackage[caption=false]{subfig}
\usepackage{xcolor}
\usepackage{enumitem}
\usepackage[normalem]{ulem}  
\usepackage{mathrsfs}

\newcommand{\postsubmissionAdd}[1]{#1}
\newcommand{\postsubmissionStrike}[1]{}

\newcommand{\minerva}{MINERvA}

\newcommand{\tune}{\minerva~Tune v1}

\newcommand{\nubar}{\bar{\nu}}
\newcommand{\enumax}{E_\nu^{\mathrm max}}
\newcommand{\etsqthresh}{{\mathscr{F}}\left( E_\mu,\theta_\mu\right) }

\begin{document}
\title{Constraining the NuMI neutrino flux using inverse muon decay reactions in MINERvA
}


\newcommand{\Rutgers}{Rutgers, The State University of New Jersey, Piscataway, New Jersey 08854, USA}
\newcommand{\Hampton}{Hampton University, Dept. of Physics, Hampton, VA 23668, USA}
\newcommand{\Dortmund}{Institute of Physics, Dortmund University, 44221, Germany }
\newcommand{\Otterbein}{Department of Physics, Otterbein University, 1 South Grove Street, Westerville, OH, 43081 USA}
\newcommand{\JMU}{James Madison University, Harrisonburg, Virginia 22807, USA}
\newcommand{\Florida}{University of Florida, Department of Physics, Gainesville, FL 32611}
\newcommand{\UCIrvine}{Department of Physics and Astronomy, University of California, Irvine, Irvine, California 92697-4575, USA}
\newcommand{\CBPF}{Centro Brasileiro de Pesquisas F\'{i}sicas, Rua Dr. Xavier Sigaud 150, Urca, Rio de Janeiro, Rio de Janeiro, 22290-180, Brazil}
\newcommand{\PUCP}{Secci\'{o}n F\'{i}sica, Departamento de Ciencias, Pontificia Universidad Cat\'{o}lica del Per\'{u}, Apartado 1761, Lima, Per\'{u}}
\newcommand{\INRM}{Institute for Nuclear Research of the Russian Academy of Sciences, 117312 Moscow, Russia}
\newcommand{\Jlab}{Jefferson Lab, 12000 Jefferson Avenue, Newport News, VA 23606, USA}
\newcommand{\Pittsburgh}{Department of Physics and Astronomy, University of Pittsburgh, Pittsburgh, Pennsylvania 15260, USA}
\newcommand{\Guanajuato}{Campus Le\'{o}n y Campus Guanajuato, Universidad de Guanajuato, Lascurain de Retana No. 5, Colonia Centro, Guanajuato 36000, Guanajuato M\'{e}xico.}
\newcommand{\Athens}{Department of Physics, University of Athens, GR-15771 Athens, Greece}
\newcommand{\Tufts}{Physics Department, Tufts University, Medford, Massachusetts 02155, USA}
\newcommand{\WM}{Department of Physics, William \& Mary, Williamsburg, Virginia 23187, USA}
\newcommand{\FNAL}{Fermi National Accelerator Laboratory, Batavia, Illinois 60510, USA}
\newcommand{\Purdue}{Department of Chemistry and Physics, Purdue University Calumet, Hammond, Indiana 46323, USA}
\newcommand{\MCLA}{Massachusetts College of Liberal Arts, 375 Church Street, North Adams, MA 01247}
\newcommand{\UMD}{Department of Physics, University of Minnesota -- Duluth, Duluth, Minnesota 55812, USA}
\newcommand{\Northwestern}{Northwestern University, Evanston, Illinois 60208}
\newcommand{\UNI}{Facultad de Ciencias, Universidad Nacional de Ingenier\'{i}a, Apartado 31139, Lima, Per\'{u}}
\newcommand{\Rochester}{University of Rochester, Rochester, New York 14627 USA}
\newcommand{\Austin}{Department of Physics, University of Texas, 1 University Station, Austin, Texas 78712, USA}
\newcommand{\USM}{Departamento de F\'{i}sica, Universidad T\'{e}cnica Federico Santa Mar\'{i}a, Avenida Espa\~{n}a 1680 Casilla 110-V, Valpara\'{i}so, Chile}
\newcommand{\Geneva}{University of Geneva, 1211 Geneva 4, Switzerland}
\newcommand{\Chicago}{Enrico Fermi Institute, University of Chicago, Chicago, IL 60637 USA}
\newcommand{\hired}{}
\newcommand{\OregonState}{Department of Physics, Oregon State University, Corvallis, Oregon 97331, USA}
\newcommand{\oxford}{Oxford University, Department of Physics, Oxford, OX1 3PJ United Kingdom}
\newcommand{\umiss}{University of Mississippi, Oxford, Mississippi 38677, USA}
\newcommand{\upenn}{Department of Physics and Astronomy, University of Pennsylvania, Philadelphia, PA 19104}
\newcommand{\AMU}{AMU Campus, Aligarh, Uttar Pradesh 202001, India}
\newcommand{\wroclaw}{University of Wroclaw, plac Uniwersytecki 1, 50-137 Wroa\l{}aw, Poland}
\newcommand{\Mohali}{Department of Physical Sciences, IISER Mohali, Knowledge City, SAS Nagar, Mohali - 140306, Punjab, India}
\newcommand{\CINVESTAV}{Departamento de Fisica Col. San Pedro Zacatenco, 07360 Mexico, DF, Av. Instituto PolitÃ©cnico Nacional, Mexico}
\newcommand{\york}{York University, Department of Physics and Astronomy, Toronto, Ontario, M3J 1P3 Canada}
\newcommand{\ND}{Department of Physics, University of Notre Dame, Notre Dame, Indiana 46556, USA}
\newcommand{\ICL}{The Blackett Laboratory,  Imperial College London,  London SW7 2BW, United Kingdom}

\newcommand{\mateusfcarneiroThanks}{Now at Brookhaven National Laboratory}
\newcommand{\finerThanks}{Now at Los Alamos National Laboratory}
\newcommand{\bamThanks}{Now at University of Minnesota}

\author{D.~Ruterbories}                   \affiliation{\Rochester}
\author{Z.~Ahmad~Dar}                    \affiliation{\WM}  \affiliation{\AMU}
\author{F.~Akbar}                         \affiliation{\AMU}
\author{M.~V.~Ascencio}                   \affiliation{\PUCP}
\author{A.~Bashyal}                       \affiliation{\OregonState}
\author{A.~Bercellie}                     \affiliation{\Rochester}
\author{M.~Betancourt}                    \affiliation{\FNAL}
\author{A.~Bodek}                         \affiliation{\Rochester}
\author{J.~L.~Bonilla}                    \affiliation{\Guanajuato}
\author{A.~Bravar}                        \affiliation{\Geneva}
\author{H.~Budd}                          \affiliation{\Rochester}
\author{G.~Caceres}                       \affiliation{\CBPF}
\author{T.~Cai}                           \affiliation{\Rochester}
\author{M.F.~Carneiro}\thanks{\mateusfcarneiroThanks}  \affiliation{\OregonState}  \affiliation{\CBPF}
\author{G.A.~D\'{i}az~}                   \affiliation{\Rochester}
\author{H.~da~Motta}                      \affiliation{\CBPF}
\author{J.~Felix}                         \affiliation{\Guanajuato}
\author{L.~Fields}                        \affiliation{\FNAL}
\author{A.~Filkins}                       \affiliation{\WM}
\author{R.~Fine}\thanks{\finerThanks}     \affiliation{\Rochester}
\author{A.M.~Gago}                        \affiliation{\PUCP}
\author{H.~Gallagher}                     \affiliation{\Tufts}
\author{A.~Ghosh}                         \affiliation{\USM}  \affiliation{\CBPF}
\author{R.~Gran}                          \affiliation{\UMD}
\author{D.A.~Harris}                      \affiliation{\york}  \affiliation{\FNAL}
\author{S.~Henry}                         \affiliation{\Rochester}
\author{D.~Jena}                          \affiliation{\FNAL}
\author{S.~Jena}                          \affiliation{\Mohali}
\author{J.~Kleykamp}                      \affiliation{\Rochester}
\author{M.~Kordosky}                      \affiliation{\WM}
\author{D.~Last}                          \affiliation{\upenn}
\author{T.~Le}                            \affiliation{\Tufts}  \affiliation{\Rutgers}
\author{A.~Lozano}                        \affiliation{\CBPF}
\author{X.-G.~Lu}                         \affiliation{\oxford}
\author{E.~Maher}
\affiliation{\MCLA}
\author{S.~Manly}                         \affiliation{\Rochester}
\author{W.A.~Mann}                        \affiliation{\Tufts}
\author{C.~Mauger}                        \affiliation{\upenn}
\author{K.S.~McFarland}                   \affiliation{\Rochester}
\author{A.M.~McGowan}                     \affiliation{\Rochester}
\author{B.~Messerly}\thanks{\bamThanks}   \affiliation{\Pittsburgh}
\author{J.~Miller}                        \affiliation{\USM}
\author{J.G.~Morf\'{i}n}                  \affiliation{\FNAL}
\author{D.~Naples}                        \affiliation{\Pittsburgh}
\author{J.K.~Nelson}                      \affiliation{\WM}
\author{C.~Nguyen}                        \affiliation{\Florida}
\author{A.~Norrick}                       \affiliation{\WM}
\author{A.~Olivier}                       \affiliation{\Rochester}
\author{V.~Paolone}                       \affiliation{\Pittsburgh}
\author{G.N.~Perdue}                      \affiliation{\FNAL}  \affiliation{\Rochester}
\author{K.-J.~Plows}                      \affiliation{\oxford}
\author{M.A.~Ram\'{i}rez}                 \affiliation{\upenn}  \affiliation{\Guanajuato}
\author{H.~Ray}                           \affiliation{\Florida}
\author{H.~Schellman}                     \affiliation{\OregonState}
\author{C.J.~Solano~Salinas}              \affiliation{\UNI}
\author{H.~Su}                            \affiliation{\Pittsburgh}
\author{M.~Sultana}                       \affiliation{\Rochester}
\author{V.S.~Syrotenko}                   \affiliation{\Tufts}
\author{E.~Valencia}                      \affiliation{\WM}  \affiliation{\Guanajuato}
\author{N.H.~Vaughan}                     \affiliation{\OregonState}
\author{A.V.~Waldron}                     \affiliation{\ICL}
\author{B.~Yaeggy}                        \affiliation{\USM}
\author{K.~Yang}                          \affiliation{\oxford}
\author{L.~Zazueta}                       \affiliation{\WM}

\collaboration{The MINER$\nu$A Collaboration}\ \noaffiliation

\date{\today}
\begin{abstract}
Inverse muon decay, $\nu_\mu e^-\to\mu^-\nu_e$, is a reaction whose cross-section can be predicted with very small uncertainties.  It has a neutrino energy threshold of $\approx 11$~GeV and can be used to constrain the high-energy part of the flux in the NuMI neutrino beam.  This reaction is the dominant source of events which only contain high-energy muons nearly parallel to the direction of the neutrino beam. 
We have isolated a sample of hundreds of such events in neutrino and anti-neutrino enhanced beams, and have constrained the predicted high-energy flux. 
\end{abstract}

\maketitle

\section{Introduction}
\label{sec:Introduction}
Neutrino oscillation experiments~\cite{Abe:2015awa,Adamson:2016tbq,Acciarri:2015uup,Abe:2011ts} depend on measurements of neutrino interactions at a near detector as a companion measurement that probes the flux and neutrino interaction cross sections that affect the experiment.  However, there are significant uncertainties both in the cross sections for neutrino interactions and in the reconstruction of neutrino energies of most reactions observed at near detectors.  These uncertainties make it difficult to use only measurements at a near detector to measure the neutrino flux and separate it from the effects of neutrino interactions.

One partial solution to this problem is to measure scattering of neutrinos from atomic electrons.  Such scattering is accurately predicted in the Standard Model, with uncertainties of a per cent or less primarily due to hadronic effects in radiative corrections~\cite{Tomalak:2019ibg}.  These reactions then provide a measurement of the flux which is independent of interaction uncertainties and can help to break degeneracies between those interaction uncertainties and uncertainties in predictions of the flux.  This technique has been demonstrated by the MINERvA experiment in $\nu e^-\to\nu e^-$ scattering~\cite{Park:2015eqa,Valencia:2019mkf} and has been studied for application in the future DUNE~\cite{Acciarri:2015uup} experiment~\cite{Marshall:2019vdy}.

Another neutrino-electron scattering reaction is inverse muon decay (IMD), $\nu_\mu e^-\to\nu_e\mu^-$.  The IMD process has a threshold energy of $E_{\mathrm{min}}=\frac{m_{\mu}^{2}-m_e^2}{2m_{e}} \approx 11$ GeV, and
a total cross section given at tree level by~\cite{Bardin:1986dk}
\begin{equation}
\sigma=\frac{(s-m_\mu^2)^2 G_{\mathrm{F}}^2}{s\pi}+{\cal O}\left( \frac{m_e^2G_{\mathrm{F}}}{s}\right) ,
\label{eqn:IMD}
\end{equation}

\noindent
where $m_{\mu,e}$ are the masses of the muon and electron, $G_F$ is the Fermi constant, and the relativistic invariant quantity, $s$, is the square of the center-of-mass scattering energy. When $E_\nu$ is measured in the lab frame, $s= 2E_\nu m_e+m_e^2$. 
The spectrum of muons emitted for a fixed neutrino energy in the lab frame, $E_\nu$, is approximately uniform with limits between $E_{\mathrm{min}}$ and $E_\nu$, with small corrections to the uniformity and the kinematic limits of order $m_e/E_\nu$ and $m_e$, respectively.  Radiative corrections to the process have been calculated, and these decrease the tree level prediction above by several percent, with the largest decreases at the lowest neutrino energies and the kinematic limits~\cite{Bardin:1986dk}.
The kinematics of IMD require 
\begin{equation}
    E_\mu\sin\theta_\mu^2=2 m_e (1-y) \left( 1+{\cal O}\left(\frac{m_e}{E_\mu}\right)+{\cal O}\left(\frac{m_\mu}{E_\mu}\right)^2\right), \label{eqn:muonAngle}
\end{equation}
where $y\equiv E_\mu/E_\nu$ and $\theta_\mu$ is the muon angle with respect to the incoming neutrino direction. Practically, this means the muon will be very close in direction to the incoming neutrino.  There is a related inverse muon decay process, $\nubar_e\to e^-\to\mu^- \nubar_\mu$, with identical kinematics and a practically indistinguishable final state.  In our experiment, the number of $\nubar_e$ above threshold is at most a few percent of the number of $\nu_\mu$ above threshold, so this contribution is unimportant.

\begin{figure}[tp]
    \centering
    \includegraphics[width=0.95\linewidth]{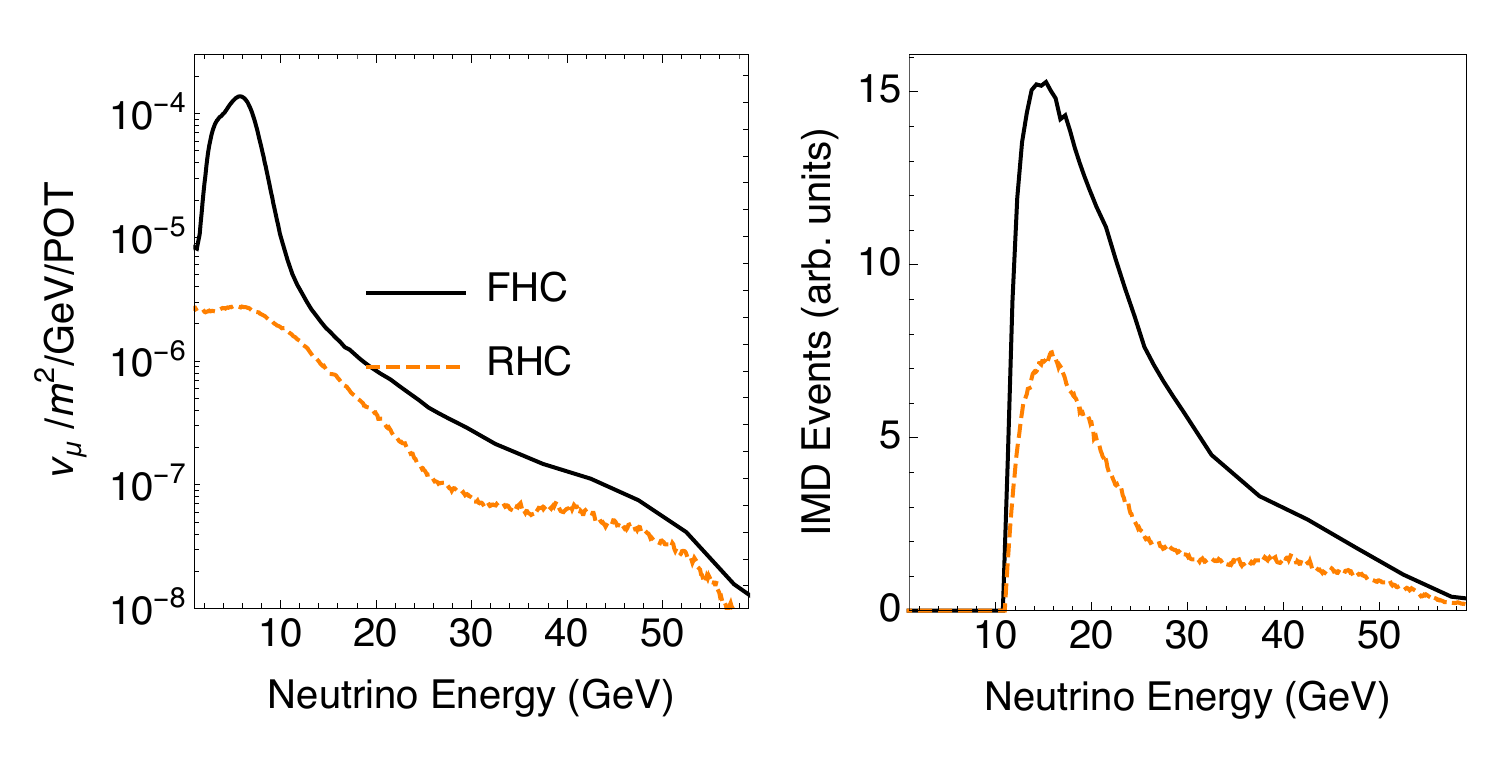}
    \caption{Predicted muon neutrino flux (left) and IMD rate (right) in the FHC and RHC NuMI beams at \minerva.}
    \label{fig:flux}
\end{figure}

For the purposes of constraining neutrino flux, IMD is only sensitive to a single neutrino type in the beam, muon neutrinos, and is only initiated by neutrinos above the threshold.  From just the spectrum of muons alone, there is only a weak correlation between muon energy and neutrino energy.  Therefore, the number of IMD events measures some weighted integral of $\nu_\mu$ over the reaction threshold.   For the NuMI neutrino beam, whose neutrino-dominated (``forward horn current'' or FHC) and anti-neutrino-dominated ``reverse horn current'' or RHC) fluxes are shown in Fig.~\ref{fig:flux}, the focusing peak is below the threshold so IMD is sensitive only to the energies greater than the focusing peak, the ``high-energy tail'', of the beam. This tail has a large contribution from neutrinos which are unfocused or under-focused by the beam optics~\cite{Anderson:1998zza,Adamson:2015dkw}.

Backgrounds to the measurement come almost entirely from high-energy neutrino $\nu_\mu$ quasielastic scattering on bound neutrons in nuclei, with small contributions from multi-nucleon and inelastic processes.  Background models described below will be improved with constraints from ``sideband'' samples at lower $E_\mu$ and higher $\theta_\mu$ than the IMD signal.

\section{The MINERvA detector and simulation}

\label{sec:Experiment}
The \minerva~experiment employs a fine-grained tracking detector for recording neutrino interactions produced by the NuMI beamline at Fermilab~\cite{Adamson:2015dkw,Aliaga:2016oaz}. 
Neutrinos are created by directing 120~GeV protons from the Main Injector onto a graphite target. 
The resulting charged pions and kaons are focused by two magnetic horns. 
Choice of the polarity of the current in the magnetic horns gives either the FHC or RHC beams, as defined above, and this analysis uses data from both beams.
Approximately 97$\%$ of the muon neutrinos that reach the \minerva~detector are produced by pion decay; the remainder are the result of kaon decay~\cite{Adamson:2015dkw,Aliaga:2016oaz}.  
At the largest neutrino energies, the fraction of neutrinos from kaons increases.  For neutrinos produced from the highest energy pions and kaons, the focusing from the horns is generally ineffective, and so the numbers of high-energy $\nu_\mu$ in the FHC and RHC beams, 
particularly those originating from $\pi^+$ decays, 
are similar.   

The \minerva~detector~\cite{Aliaga:2013uqz} consists of 120 hexagonal modules that create an active tracking volume preceded by a set of passive nuclear targets. This result includes only those interactions in the active tracking volume with a fiducial mass of 5.48 tons. The
active target volume is surrounded by electromagnetic and hadronic calorimeters.
Each tracking module has two planes composed of triangular polystyrene scintillator strips with a 1.7 cm strip-to-strip pitch. For  three-dimensional reconstruction, planes are oriented in three different directions, 0$^{\circ}$ and $\pm$ 60$^{\circ}$ relative to the vertical axis of the detector. The downstream and side electromagnetic calorimeters consist of alternating layers of scintillator and 2~mm-thick lead planes. The downstream hadronic calorimeter consists of alternating scintillator and 2.54~cm-thick steel planes.
Multi-anode photomultiplier tubes read out the scintillator strips via wavelength-shifting fibers. The timing resolution of the readout electronics is 3.0 ns and sufficient to separate multiple interactions within a single NuMI beam spill. 

Muons that originate in MINERvA from IMD travel entirely through MINERvA into the   MINOS near detector~\cite{Michael:2008bc} located 2~m downstream of the \minerva~detector.  In MINOS, their momentum and electric charge are measured by a magnetized spectrometer composed of scintillator and iron.

This analysis uses data that correspond to $10.61\times 10^{20}$ protons on target (POT) in the FHC configuration and $11.24\times 10^{20}$ POT in the RHC configuration taken between September 2013 and February 2019.  The beam focusing configuration and target are that of the ``medium energy'' beam provided for the NOvA experiment.

A GEANT4-based simulation of the NuMI beamline is used to predict the neutrino flux. To improve the prediction, the simulation is reweighted as a function of pion kinematics to correct for differences between the GEANT4~\cite{Agostinelli:2002hh}
prediction and hadron production measurements of 158~GeV protons on carbon from the NA49 experiment~\cite{Alt:2006fr} and other relevant hadron production measurements. A description of this procedure is found in Ref.~\cite{Aliaga:2016oaz}. The {\em in situ} measurement of neutrino scattering off atomic electrons  described in Ref.~\cite{Valencia:2019mkf}, is not used in this analysis to constrain the flux prediction.  This measurement and the neutrino-electron elastic scattering measurements give independent constraints which may be combined.

\begin{figure}[tp]
    \centering
    \includegraphics[width=0.99\linewidth]{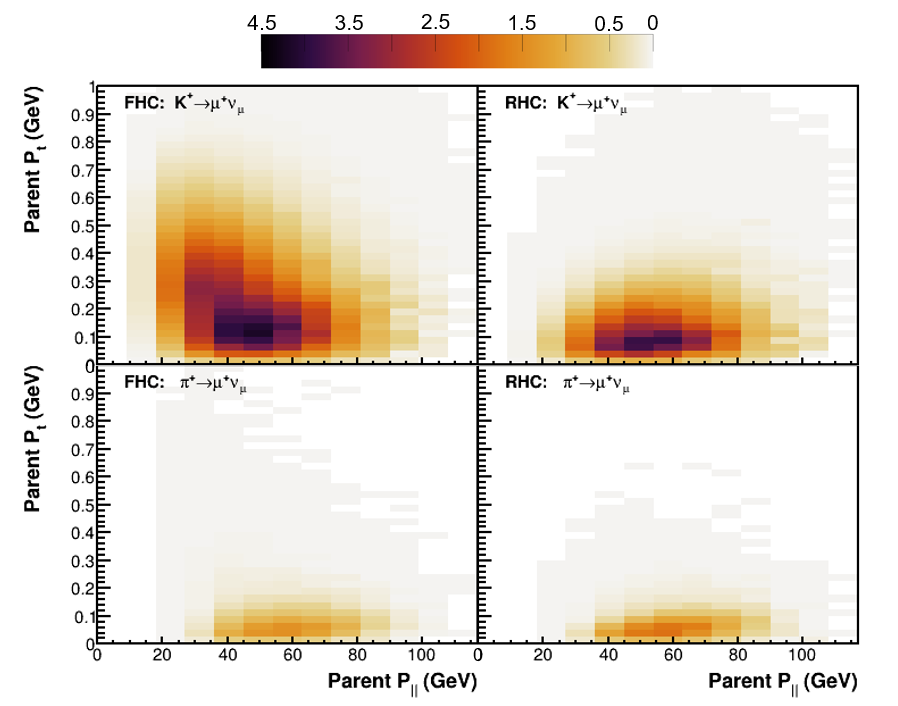}
    \caption{The predicted number of IMD events in the MINERvA detector fiducial volume in bins of the parent $\pi^+$ or $K^+$ longitudinal and transverse momentum. Neutrinos from kaon parents are in the top row while pions are on the bottom row. The FHC beam is shown in the left column and RHC in the right column.}
    \label{fig:fluxParents}
\end{figure}

The kinematics of the parent mesons for IMD neutrinos in both FHC and RHC, as predicted by the simulation, are shown in Fig.~\ref{fig:fluxParents}.  This simulation predicts two dominant populations of mesons that produce neutrinos with sufficient energy to contribute to the inverse muon decay signal. The fraction of neutrinos from $\pi^+$ decay is 9 (17)\% percent in the FHC (RHC) beam. The first population consists of $K^+$ at moderate longitudinal momentum, $p_{||}\stackrel{>}{\sim}30$~GeV, and with a range of magnitudes of momenta transverse to the proton beam direction, $p_T$.  The second population consists of $\pi^+$ and $K^+$ at higher $p_{||}$, $\stackrel{>}{\sim}40$~GeV and $p_T\stackrel{<}{\sim}0.15$~GeV.  The second population is dominated by mesons which are underfocused or entirely unfocused by the horns and are common to the FHC and RHC predictions, whereas the first is a unique contribution in the FHC beam since these high-$p_T$ 
$K^+$ are defocused in the RHC beam.

Neutrino interactions are simulated using the GENIE neutrino event generator~\cite{Andreopoulos:2009rq} version 2.12.6. Quasi-elastic (1p1h) interactions are simulated using the Llewellyn-Smith formalism~\cite{LlewellynSmith:1971zm} with the vector form factors modeled using the BBBA05 model~\cite{Bradford:2006yz}. The axial vector form factor uses the dipole form with an axial mass of $M_A=0.99$ GeV/c$^2$. Resonance production is simulated using the Rein-Sehgal model~\cite{Rein:1980wg} with an axial mass of $M_A^{RES}=1.12$ GeV/c$^2$. 
Higher invariant mass interactions, including  
Deep Inelastic Scattering (DIS), are simulated using a leading-order pQCD model with the Bodek-Yang prescription~\cite{Bodek:2004pc} for the modification at low square of the momentum transfer, $Q^2$.

A relativistic Fermi gas model~\cite{Smith:1972xh} with an additional Bodek-Ritchie high momentum tail~\cite{Bodek:1981wr}
 is used to describe the nuclear environment. The maximum momentum for Fermi motion is assumed to be $k_F=0.221$ GeV/c.  GENIE models intranuclear rescattering, or final state interactions (FSI), of the produced hadrons using the INTRANUKE-hA package~\cite{Dytman:2007zz}.

To better describe \minerva~data, a variety of modifications to the interaction model are made. To better simulate quasielastic events, the cross section is modified as a function of energy and three momentum transfer 
based on the random phase approximation (RPA) part of the Valencia model~\cite{Nieves:2004wx,Gran:2017psn} appropriate for a Fermi gas~\cite{Martini:2016eec,Nieves:2017lij}. Multi-nucleon scattering (2p2h) is simulated by the same Valencia model~\cite{Nieves:2011pp,Gran:2013kda,Schwehr:2016pvn}, but the cross section is increased in specific regions of 
energy and three momentum transfer based on fits to \minerva~ data~\cite{Rodrigues:2015hik} in a lower energy beam configuration. Integrated over all phase space, the rate of 2p2h is increased by 50\% over the nominal prediction. Based on fits done in Ref.~\cite{Rodrigues:2016xjj}, we decrease the non-resonant pion production by 43\% and reduce the uncertainty compared to the base GENIE model uncertainties. This modified version of the simulation is referred to later in this paper as \tune.

The response of the \minerva~detector is simulated using GEANT4~\cite{Agostinelli:2002hh} version 4.9.3p6 with the QGSP\_BERT physics list. The optical and electronics performance is also simulated. Through-going muons are used to set the absolute energy scale of minimum ionizing energy depositions by requiring the average and RMS of energy deposits match between data and simulation as a function of time. A full description is found in Ref.~\cite{Aliaga:2013uqz}. Measurements using a charged particle test beam~\cite{Aliaga:2015aqe} and a scaled-down version of the \minerva~detector set the absolute energy response to charged hadrons. The effects of accidental activity are simulated by overlaying hits in both \minerva~and MINOS from data corresponding to random beam spills appropriate to the time periods in the simulation. 

\section{Selection of Inverse Muon Decay Events}
\label{sec:selection}

A charged-current $\nu_\mu$ event is selected by matching a reconstructed muon track in MINERvA with a momentum and charged analyzed muon track in MINOS.
The approximately 252(132) event IMD sample, prior to $E_{\mu}$ and $\theta_\mu$ selections, in FHC(RHC) is a small subsample, approximately $0.006\%$
of the inclusive $\nu_\mu$ charged-current sample with a reconstructed neutrino interaction point in the tracking fiducial volume. 
For the high-energy muons in the IMD sample, the selection of $\mu^-$ using the direction of the bend in the magnetic MINOS spectrometer is 99\% efficient, and in the RHC sample where most muons are $\mu^+$, the purity for $\mu^-$ selection is over 97\%.

Since IMD only produces an energetic forward muon in the final state, the visible energy in the tracker and calorimeters is expected to come only from the muon.  By contrast, events from background reactions on nuclei almost always produce some visible recoil, including 
a recoiling target nucleon.  
The IMD selection requires visible hadronic energy in tracker and electromagnetic calorimeters be less than $80$~MeV and visible energy 
within $150$~mm of the neutrino interaction point be less than $10$~MeV.  This effectively removes most events with low-energy protons or pions in the final state while retaining all but 4\% of the IMD events.

In addition, the reconstructed muon must be a $\mu^-$ with total energy greater than $10$~GeV, a threshold below the kinematic threshold of the IMD process due to the $\approx 11\%$ fractional energy resolution in MINOS.  Negatively charged muon candidates are also required to have reconstructed energy below $50$~GeV, a point beyond which the 
charge of the muon cannot be reliably measured.  From Eq.~\ref{eqn:muonAngle}, the kinematics of neutrino scattering from atomic electrons requires that $E_\mu\sin^2\theta_\mu = 2m_e\left(1-\frac{E_\mu}{E_\nu}\right)$, where $\theta_\mu$ is the scattering angle with respect to the initial neutrino direction. 
In a given event, we measure $E_\mu$ and $\theta_\mu$, but we do not know $E_\nu$ {\em a priori}, nor do we measure $\theta_\mu$ with sufficient precision to extract $E_\nu$ from the relationship above.  However, we have a minimum $E_\mu$ for IMD events, and we have a maximum relevant $E_\nu$ set by our flux which falls steeply with energy as shown in Fig.~\ref{fig:flux}.  These facts together imply that $1-\frac{E_\mu}{E_\nu}$ will typically be a number significantly less than $1$, thus allowing us to place a tighter selection on $E_\mu\sin^2\theta_\mu$ than the maximum of $2m_e$ which is reached in the limit of $\frac{E_\mu}{E_\nu}\to 0$.  We define $\enumax\equiv 35$~GeV, and form
\begin{equation}
\etsqthresh \equiv \frac{E_\mu\frac{\theta_\mu^2}{1 {\mathrm radian}^2}}{1-\frac{E_\mu}{\enumax}},
\label{eqn:etheta2thresh}
\end{equation}
\noindent where the small angle expansion has been used to set $\sin\theta_\mu\approx \frac{\theta_\mu}{1 {\mathrm radian}}$. The event selection then requires $\etsqthresh <2m_e$.  This cut will be increasingly inefficient for neutrino energies above $\enumax$, but the choice of $35$~GeV is predicted to include $98\%$ of the IMD events in the FHC beam and $75\%$ in the RHC beam before accounting for experimental resolutions.
The distribution in $\etsqthresh$ after all selections and background tuning is shown in Fig.~\ref{fig:signal_constrained_etheta2}.

\begin{figure}[tp]
    \centering
    \includegraphics[width=0.49\linewidth]{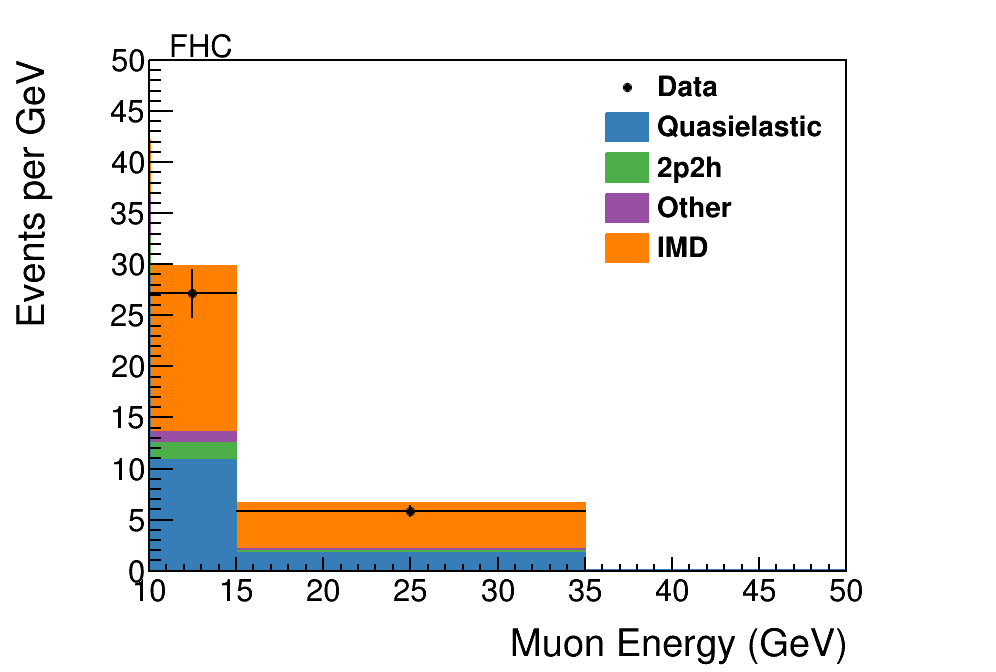}
    \includegraphics[width=0.49\linewidth]{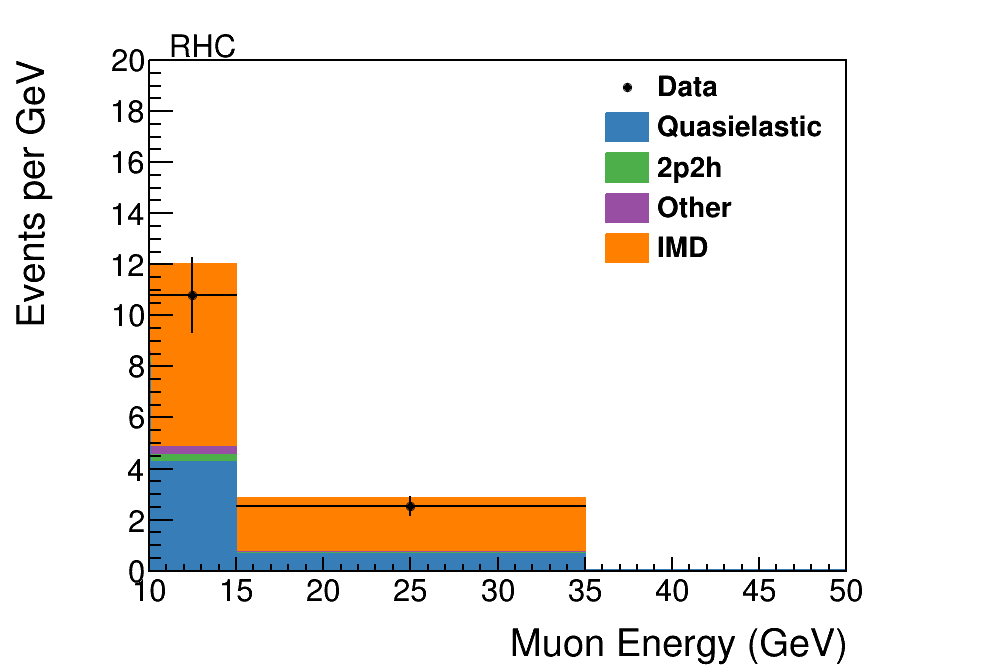}
    \caption{Selected signal channel events as a function of muon energy. The FHC sample is on the left and RHC sample is on the right. There are significantly fewer events in the RHC sample than the FHC sample, as expected.
    }
    \label{fig:selectedsignal}
\end{figure}

Figure~\ref{fig:selectedsignal} shows the expected and observed signal sample as a function of reconstructed muon energy in the FHC and RHC beams after these selections.  The expected signal and background are nearly comparable, and the backgrounds are almost entirely due to charged-current quasielastic scattering, $\nu_\mu n_{\rm bound} \to\mu^- p$, with a small fraction of events from the multi-nucleon version of this scattering, the 2p2h process described above.

\subsection{Background Constraints}
\label{subsec:Sideband}
\begin{figure}[tp]
    \centering
    \includegraphics[width=0.49\linewidth]{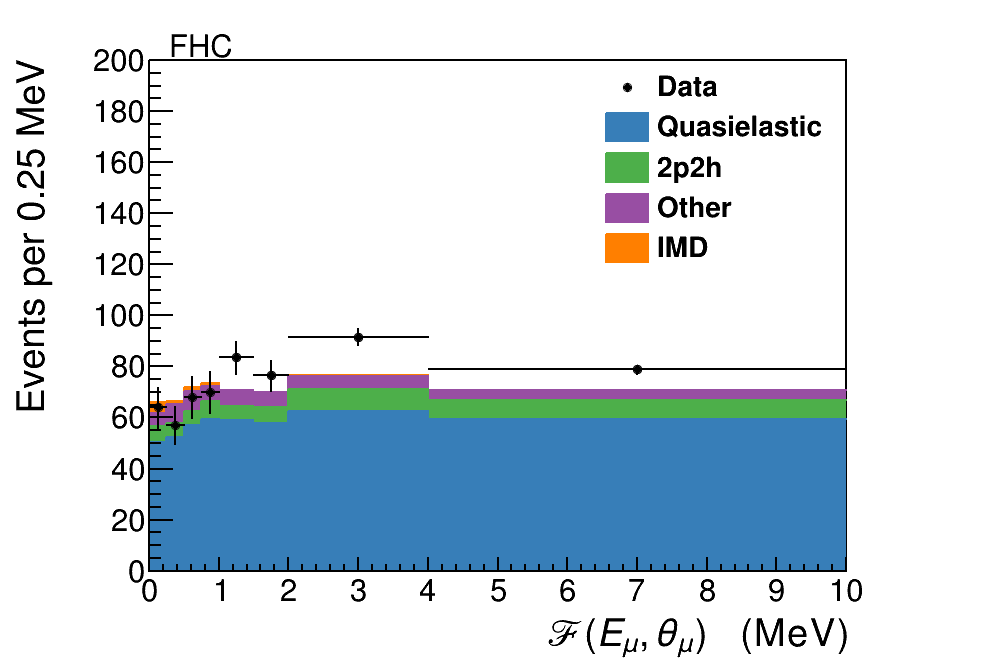}
    \includegraphics[width=0.49\linewidth]{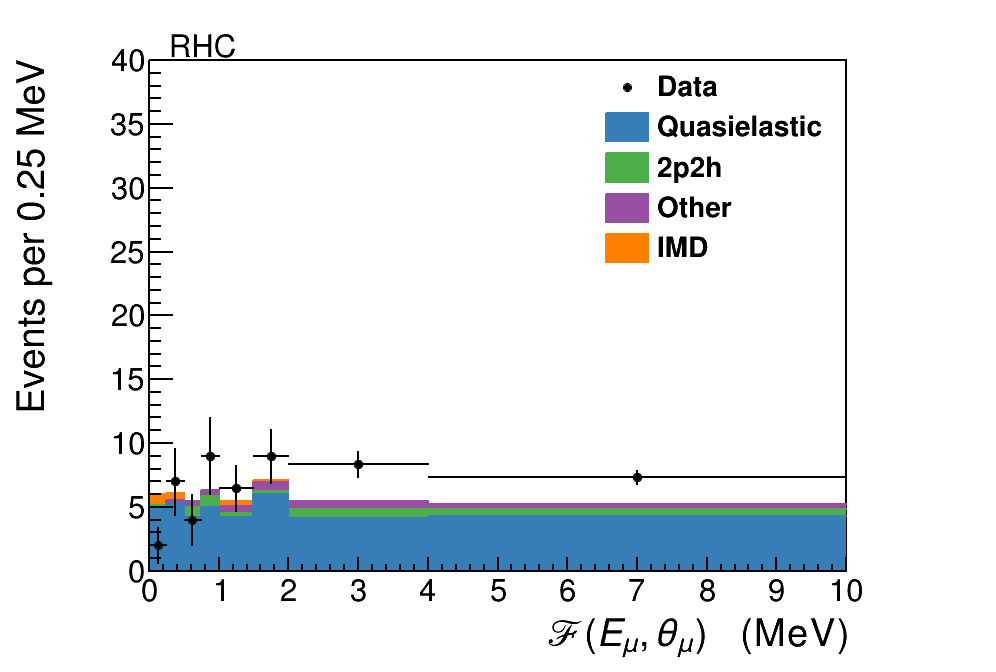}\\
    \includegraphics[width=0.49\linewidth]{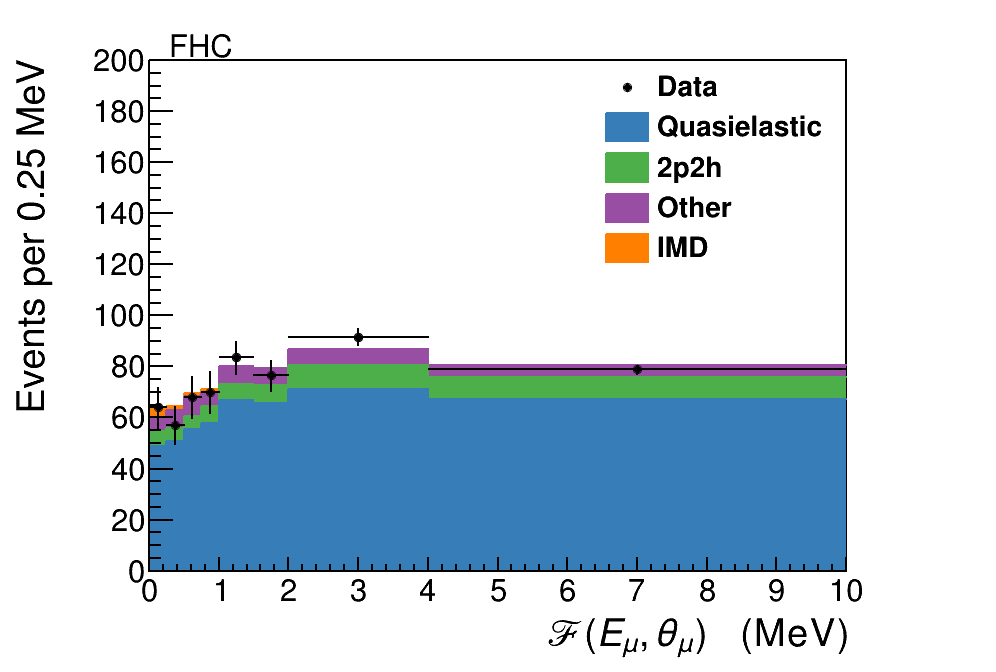}
    \includegraphics[width=0.49\linewidth]{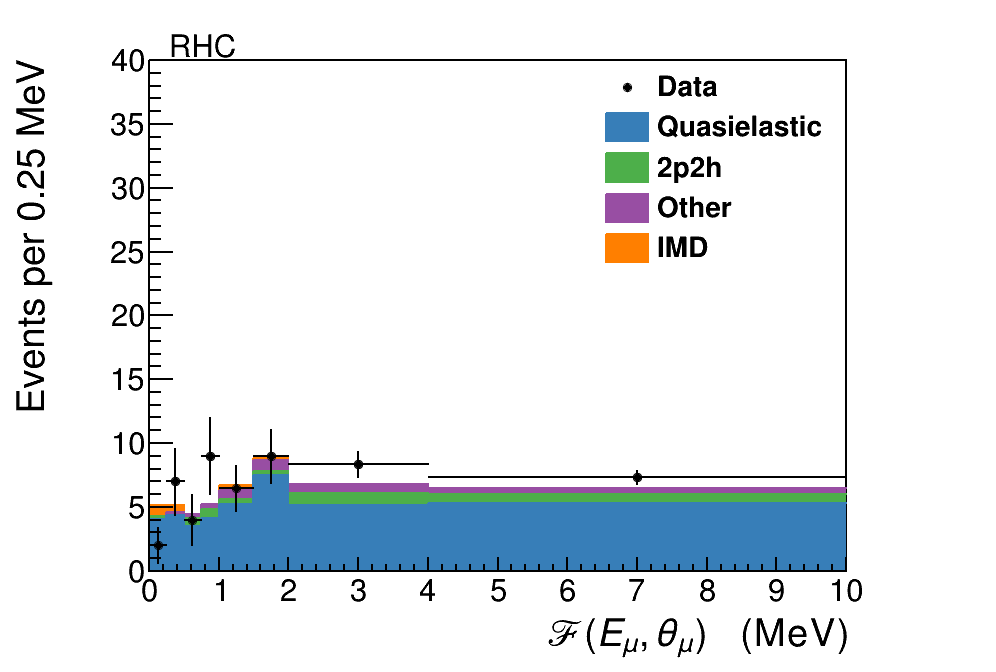}
    \caption{Sideband samples before (top) and after (bottom) the application of the fit results as a function of $\etsqthresh$, defined in Eq.~\ref{eqn:etheta2thresh}. The FHC sample is on the left, and the RHC sample is on the right. 
    }
    \label{fig:sideband}
\end{figure}

To constrain the remaining background a sideband sample is measured using events which pass the recoil and vertex energy criteria, but have muon energy between $7$ and $9$~GeV. The sample composition is almost exactly the same as the backgrounds in the signal selection, but this sample has almost no signal component.  Figure~\ref{fig:sideband} shows the sideband sample as a function of $\etsqthresh$.  As can be seen particularly with the FHC sample, there are two differences between the sideband simulation and data, both likely due to poorly modeled nuclear effects.  The first is the overall rate, which will be strongly affected by the probability that outgoing nucleons reinteract to produce neutrons in the final state, which in turn go undetected and allow the events to pass the recoil cuts.  The second is that the events at low $\etsqthresh$ are suppressed, possibly due to Pauli blocking or nuclear screening.  This sideband sample is used to divide the background into events with $\etsqthresh$ below and above $2m_e$, where the former is the one that directly enters into the background subtraction for the signal sample.  However, the absolute data and simulation differences between the sideband and the signal with $E_\mu>10$~GeV are also affected by uncertainties in the flux itself. The flux uncertainties are rapidly changing in the sample since the focusing peak is at $E_\nu\sim7$~GeV.  Therefore a second sample of events with $E_\mu>10$~GeV but with $4\leq \etsqthresh < 10$~MeV is added to provide an absolute normalization to the background prediction.  The resulting scale factors and their uncertainties are shown in Table~\ref{tab:sideband_fit}, and the corrected simulated distributions compared to the sideband data are shown in Fig~\ref{fig:sideband}.  The net effect is to increase the backgrounds compared to the prediction by approximately $15\%$ in the high $\etsqthresh$ region, but to suppress the background in the signal region of $\etsqthresh<2m_e$.

\begin{table*}[btp]
 \centering
 \begin{tabular}{l|c|c|c}
    & \multicolumn{2}{c|}{$7<E_\mu \leq 9~$ GeV and} & $E_\mu \ge 10$~GeV and \\
    Beam  &  $\etsqthresh\leq 2m_e$  &$\etsqthresh> 2m_e$   &  $4\leq \etsqthresh < 10$~MeV.  \\ \hline
     FHC & $0.97\pm 0.05$ & $1.13\pm 0.02$ & $1.16 \pm 0.04$\\ 
      RHC & $0.83\pm 0.19$ & $1.25\pm 0.06$ & $1.15\pm 0.06$
 \end{tabular}
 \caption{Background scale factors for the FHC and RHC samples.   The scale factors applied to the signal region use the $7<E_\mu<9$~GeV region to find the fraction of background events with $\etsqthresh<2m_e$, and the $E_\mu>10$~GeV, high $\etsqthresh$ region to normalize the background distributions.
 }
    \label{tab:sideband_fit}
\end{table*}

\begin{figure}[tp]
    \centering
    \includegraphics[width=0.49\linewidth]{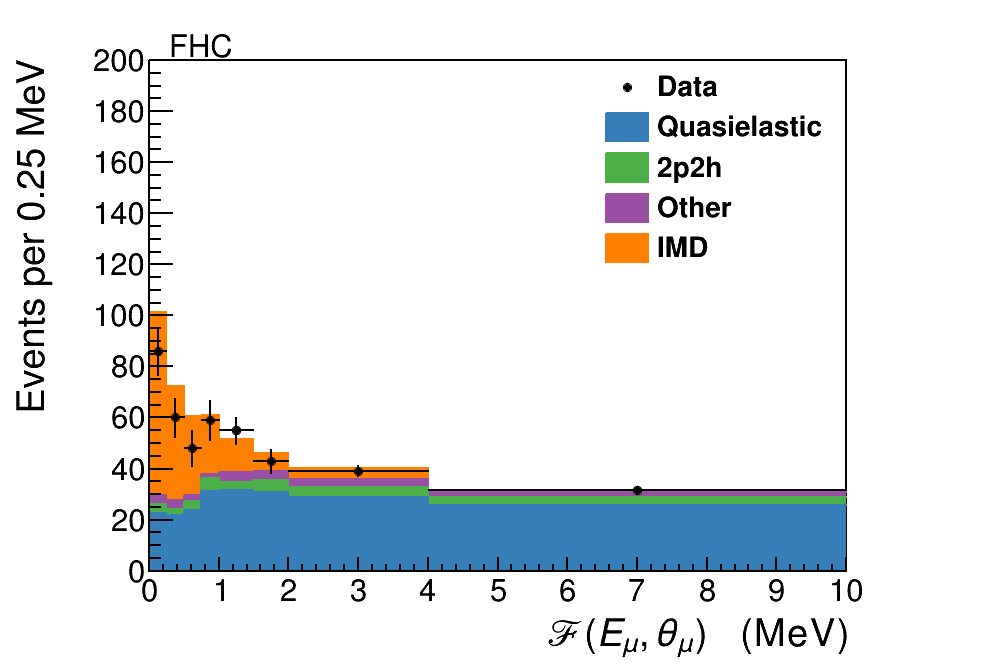}
    \includegraphics[width=0.49\linewidth]{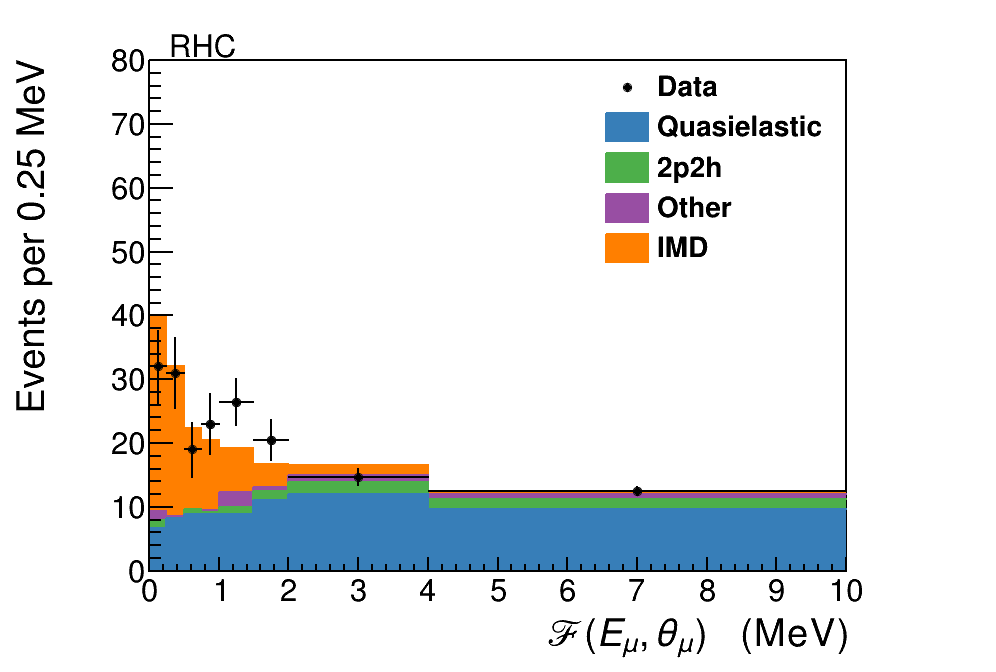}
    \caption{Selected signal samples and nearby higher $\etsqthresh$ after the application of the sideband fit results as a function of $\etsqthresh$. The FHC sample is on the left, and the RHC sample is on the right. 
    }
    \label{fig:signal_constrained_etheta2}
\end{figure}

\begin{figure}[tp]
    \centering
    \includegraphics[width=0.49\linewidth]{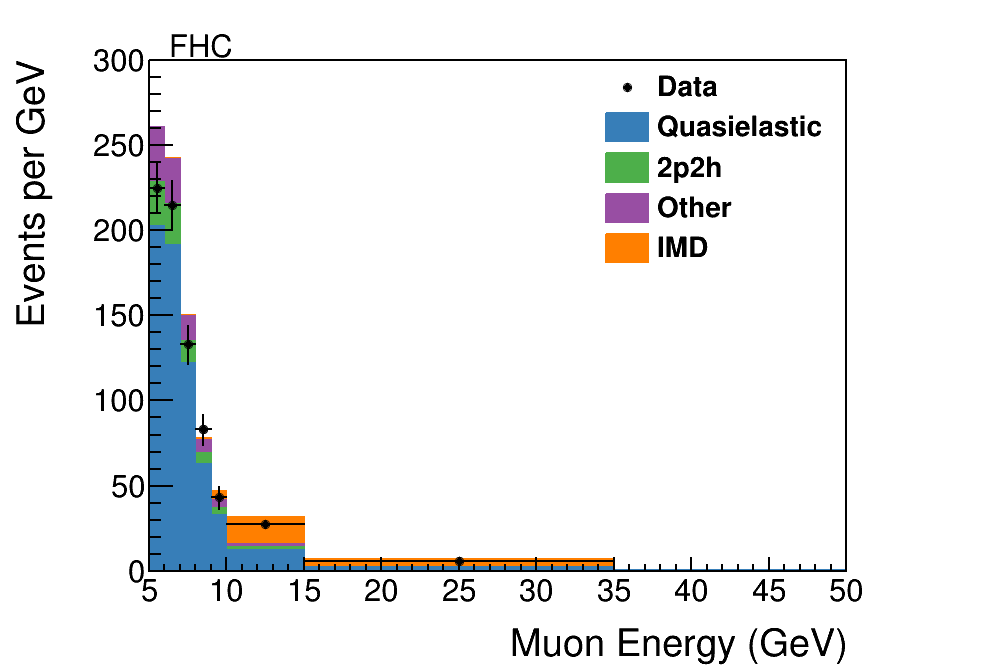}
    \includegraphics[width=0.49\linewidth]{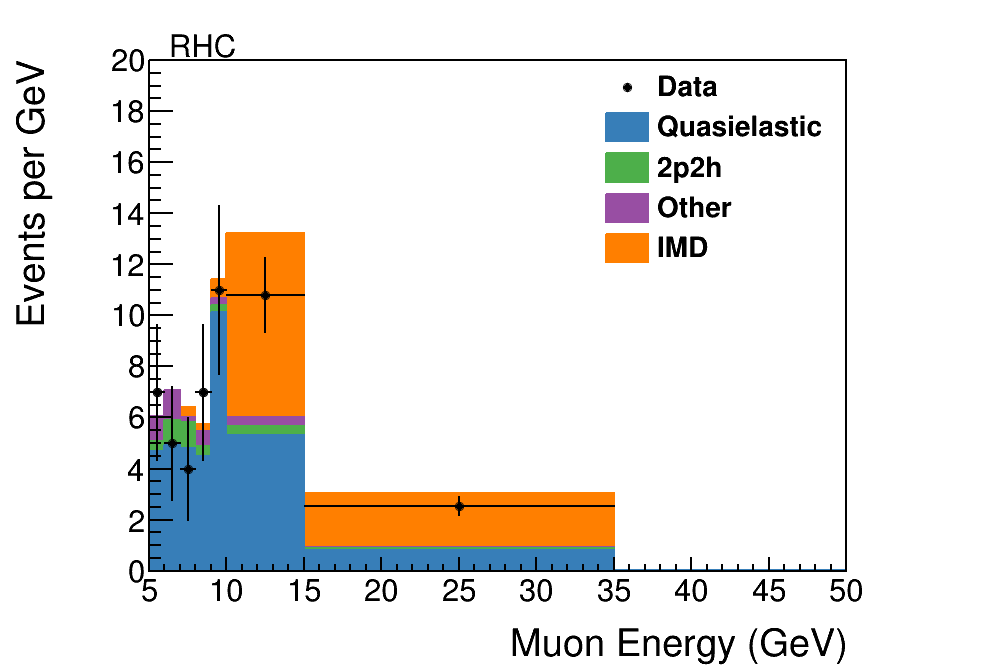}
    \caption{Selected signal samples and nearby lower $E_\mu$ events after the application of the sideband fit results as a function of muon energy. The FHC sample is on the left and RHC sample is on the right. 
    }
    \label{fig:signal_constrained_E}
\end{figure}

After the sideband fit, the scale factors are applied to the selected signal sample. The resulting distribution is shown as a function of $\etsqthresh$ 
and muon energy in Figs.~\ref{fig:signal_constrained_etheta2} and \ref{fig:signal_constrained_E},  respectively.  

\subsection{Systematic Uncertainties}

Systematic uncertainties in this analysis fall under three different categories:  flux, detector response, and neutrino interaction model uncertainties.  
The uncertainties from individual sources are evaluated by re-extracting background subtracted samples using modified simulations. The size of each modification is related to the uncertainty in each source. Neutrino interaction model uncertainty in the result is solely due to the background interactions since the signal interaction model is well known. 

The flux uncertainty is a typical leading uncertainty in neutrino cross section measurements, but since in this analysis the output is just a count of the number of events, the flux uncertainty enters only through the background constraint which is extrapolated 
from lower muon and presumably neutrino energy, to higher energy.  The resulting small uncertainties from the input flux on the background subtraction are compared with the \emph{a priori} flux when this result is applied as a flux constraint.  A related uncertainty which must be factored into the predicted number of events is the normalization uncertainty of 1.4\% from uncertainty in the number of electrons in the target, based on material assays and weight measurements of scintillator planes.  

The uncertainty in the detector response to hadrons is evaluated using shifts determined by \textit{in situ} measurements of a smaller version of the detector in a test beam~\cite{Aliaga:2015aqe}. Uncertainties in inelastic interaction cross sections for particles in the detector material are independently varied based on data-Monte Carlo differences between GEANT particle cross sections and world data on neutrons \cite{Abfalterer:2001gw,Schimmerling:1973bb,Voss:1956,Zanelli:1981zz}, pions \cite{Ashery:1981tq,Allardyce:1973ce,Wilkin:1973xd,Clough:1974qt}, and protons \cite{Menet:1971zz,Dicello:1970mx,McGill:1974zz}. 
The muon reconstruction uncertainty is dominated by uncertainty in the energy scale, which is constrained by a combination of data and simulation described in Ref.~\cite{Carneiro:2019jds} to $1.0\%$.  
The uncertainty in the matching efficiency is from imperfect modeling of the efficiency loss from accidental activity in the MINOS near detector when matching muon tracks from MINERvA to MINOS.  This last efficiency is also determined by a data-simulation comparison as a function of instantaneous neutrino beam intensity.  

The interaction model uncertainties are evaluated using the standard GENIE reweighting infrastructure with additional uncertainties from \tune. The sideband constraint reduces those uncertainties by more than a factor of two.

The final samples have 127(56) selected events in data for the FHC(RHC) configurations.  Due to the limited size of the sample, each is only reported as total number of events.  The statistical and systematic uncertainties in the background subtracted samples are shown in Tab.~\ref{tab:finalsample}.  In the both beams, the prediction is larger than the 
observed number of events as shown in Fig.~\ref{fig:universes_measurement}.  Both results are dominated by statistical uncertainty with subleading contributions from the 
uncertainties in 
the interaction cross section model and the muon reconstruction.  The $\nubar_e$ initiated reaction described above is predicted by the cross-section in Ref.~\cite{Marciano:2003eq} to be $0.5\%$($2\%$) of the signal rates above, and for convenience was treated as a background in this analysis.

\begin{table}[btp]
 \centering

\begin{tabular}{l|c|c}
     & FHC & RHC \\ \hline
    Total Uncertainty  &  \postsubmissionAdd{20.4} &	\postsubmissionAdd{11.4} \\ \hline
    Individual Uncertainties & & \\
    ~~Statistical &	\postsubmissionAdd{16.7} &	\postsubmissionAdd{11.0} \\
    ~~Background Interaction Model &	\postsubmissionAdd{6.9} &	\postsubmissionAdd{0.9} \\
    ~~Final State Interaction Model &	\postsubmissionAdd{6.9} &	\postsubmissionAdd{1.0} \\
    ~~Flux & \postsubmissionAdd{5.2} &	\postsubmissionAdd{1.8} \\
    ~~Muon Reconstruction &	\postsubmissionAdd{3.8} &	\postsubmissionAdd{1.0} \\
    ~~Others &	\postsubmissionAdd{1.9} &	\postsubmissionAdd{1.6}  \\
    \hline IMD Events in Sample & \postsubmissionAdd{127}. & \postsubmissionAdd{56.} \\ \hline
\end{tabular}
    \caption{Background subtracted sample and systematic uncertainties, in \postsubmissionAdd{the number of predicted or measured} events, on the measurement.  
    }
    \label{tab:finalsample}
\end{table}

\section{Flux constraints from IMD}
\label{sec:Results}

\begin{figure}[tp]
    \centering
    \includegraphics[width=0.75\linewidth]{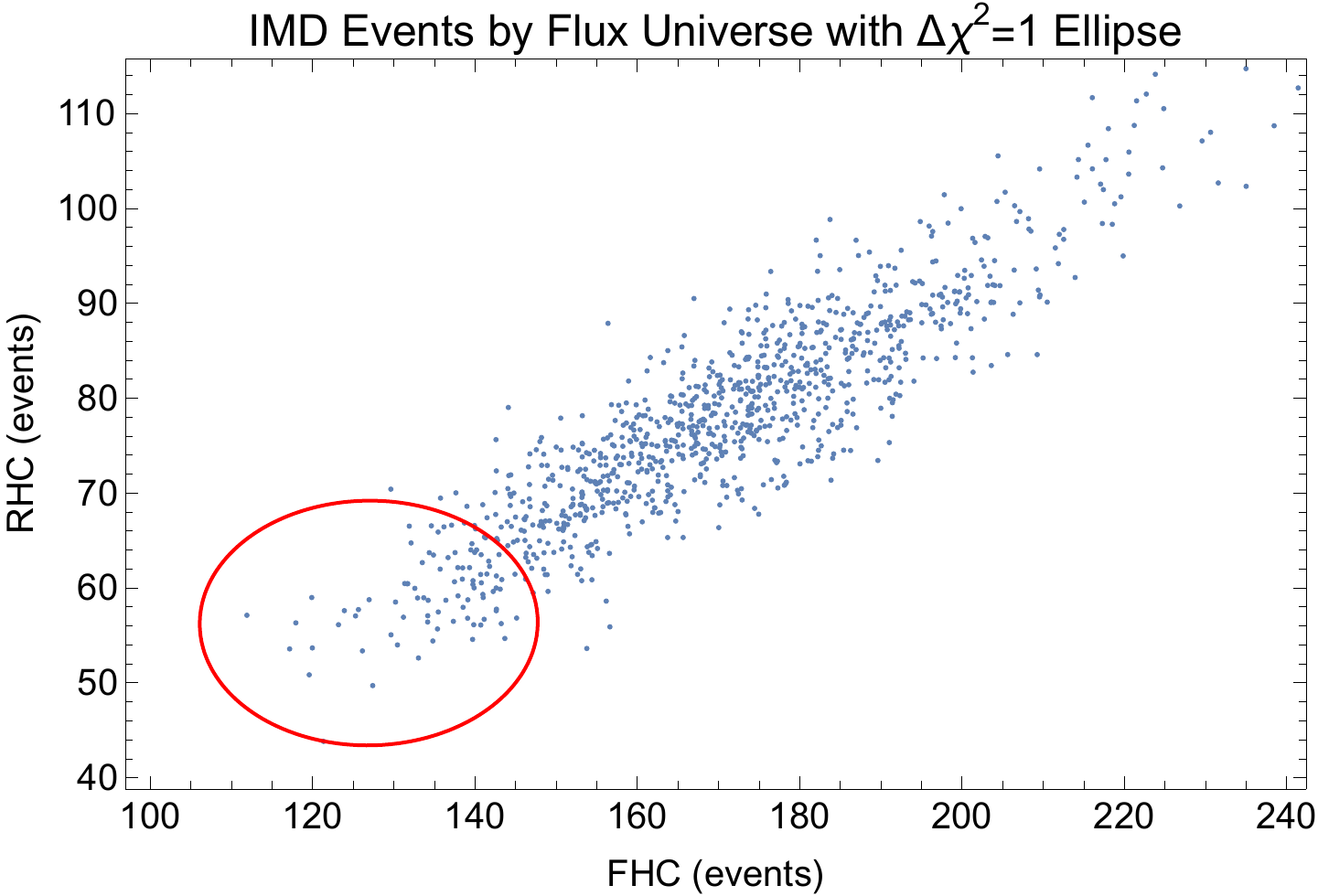}
    \caption{The predicted $N_{RHC}$ vs $N_{FHC}$ in each of 1000 flux universes, with the superimposed $\Delta\chi^2=1$ ellipse of events predictions consistent with the measurements.  
    }
    \label{fig:universes_measurement}
\end{figure}

The prediction of the MINERvA flux~\cite{Aliaga:2016oaz,Park:2015eqa,Valencia:2019mkf} described in Section~\ref{sec:Experiment} gives a nominal 
flux prediction, the ``central value'', and a series of flux ``universes'' that describe the uncertainties and covariances in those uncertainties by the Monte Carlo method.  The consistency of each flux universe, denoted by $\Phi$ with a universe index, $i$,
with the number of IMD events, $N$,   is measured by the probability of the measurement given this flux, $P(N|\Phi_i)$.  Since this measurement consists of two weakly correlated measurements in the two beams, the consistency of these measurements with a given flux universe is given by
\begin{equation}
P(\left\{ N_{FHC},N_{RHC}\right\} |\Phi^{(FHC,RHC)}_i)=\frac{1}{2\pi\sqrt{\left| V\right|}}e^\frac{-\Delta^T V^{-1}\Delta}{2},
\label{eqn:jointProb}\end{equation}
where $\Delta$ is a vector of the difference between the number of measured and predicted events in the FHC and RHC beams, and $V$ is the covariance matrix of these measurements.
Figure~\ref{fig:universes_measurement} shows the predicted number of IMD events in the FHC and RHC beams with the measurement superimposed.  It is evident from this that some flux universes are significantly less consistent with the measurement than others. 
The \emph{a priori} prediction of the flux can then be modified, according to Bayes' Theorem, by weighting the flux universes  by the probability given in Eq.~\ref{eqn:jointProb} when forming the central value prediction or the variance of the ensemble of universes.  The evident correlation between the FHC and RHC predictions is in part due to the common source of high-energy $\nu_\mu$ in the two beams of unfocused low $p_T$ parent mesons, as discussed in Sec.~\ref{sec:Experiment}.

\begin{figure}[tp]
    \centering
    \includegraphics[width=0.49\linewidth]{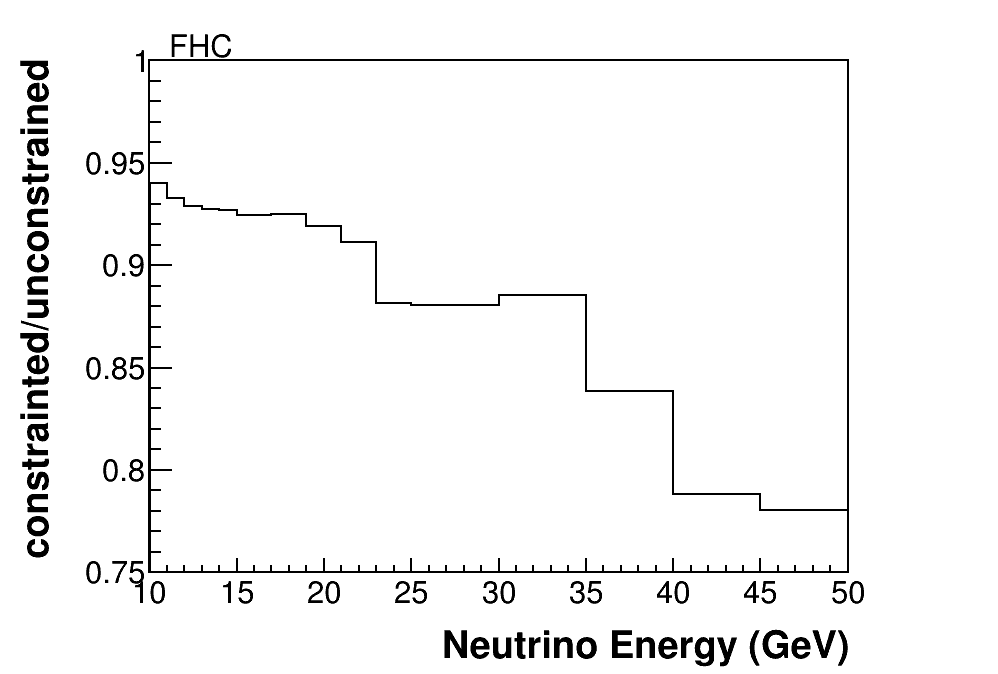}
    \includegraphics[width=0.49\linewidth]{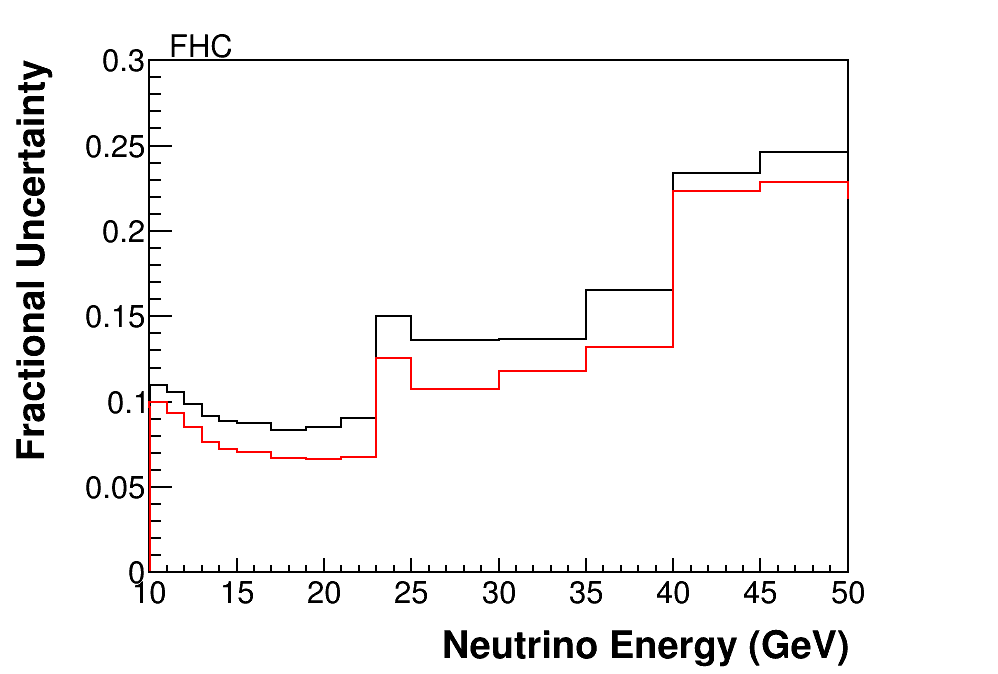}
    \includegraphics[width=0.49\linewidth]{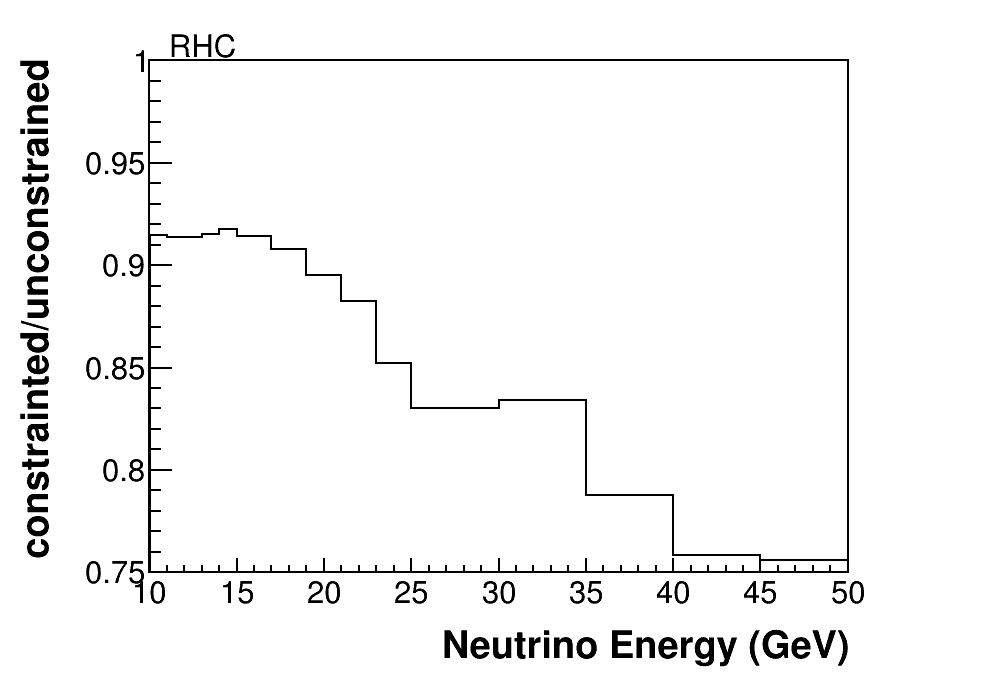}
    \includegraphics[width=0.49\linewidth]{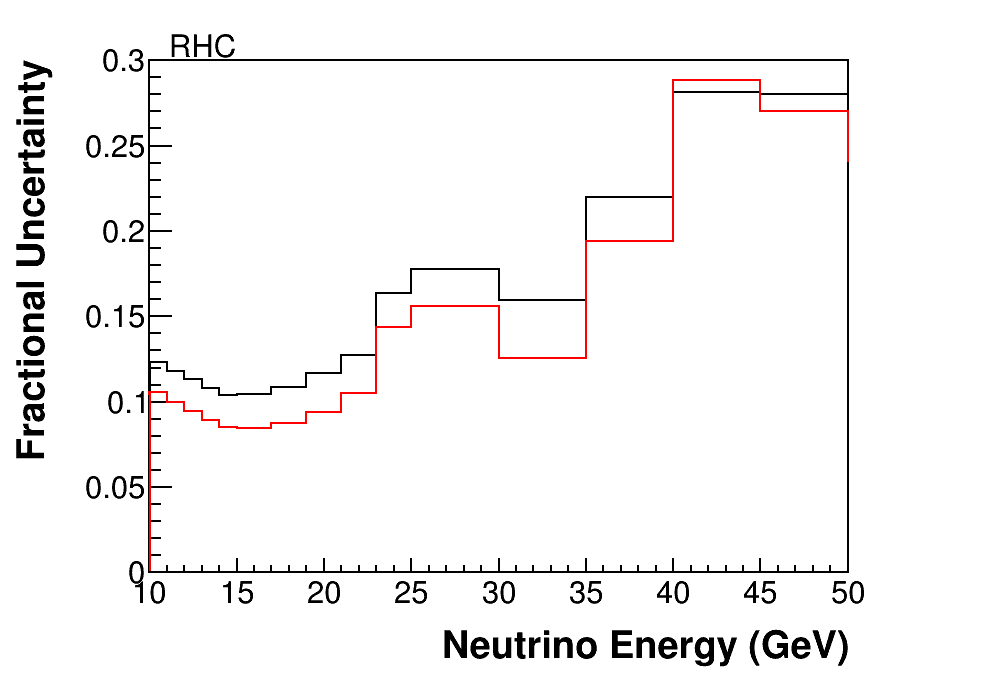}
    \caption{The FHC (top) and RHC (bottom) fluxes (left), and their uncertainties (right), before (black) and after (red) the constraint as a function of $E_\nu$ 
    }
    \label{fig:fluxPostFit}
\end{figure}

The predicted RHC and FHC fluxes 
and their uncertainties, before and after the IMD constraints are applied, are shown in Fig.~\ref{fig:fluxPostFit}.  As expected, the IMD sample provides a significant constraint for the highest energy neutrinos in the NuMI beams. 
The flux is modified and constrained at neutrino energies below the threshold of $\approx 11$~GeV because high-energy mesons that decay to make these neutrinos may also decay at larger angles with respect to the beam axis to produce lower energy neutrinos.
Therefore, even though we only measure the IMD rate above threshold, we are constraining flux universes that encode the physics that leads to that high-energy part of the neutrino spectrum.
The integral flux above $11$~GeV \postsubmissionAdd{in the FHC beam} is predicted to be $2.61\times10^{-5}\pm 2.28\times10^{-6}$ $\nu_\mu$/POT/m$^2$ before the constraint and is evaluated as $2.38\times10^{-5}\pm 1.50\times10^{-6}$ $\nu_\mu$/POT/m$^2$ after.

\section{Conclusions}
\label{sec:Conclusion}
The MINERvA experiment has successfully isolated a sample of inverse muon decay events, $\nu_\mu e^-\to\mu^- \nu_e$, and has used those events to constrain the flux of high-energy neutrinos in its beam.  The constraint provides an \emph{in situ} way to reduce uncertainties from its high-energy flux. Such a method can be applied to any accelerator neutrino beam produced by protons of energies much greater than the $11$~GeV threshold for inverse muon decay, and in particular can be used for a similar purpose in the planned DUNE experiment.

\begin{acknowledgments}

This document was prepared by members of the MINERvA Collaboration using the resources of the Fermi National Accelerator Laboratory (Fermilab), a U.S. Department of Energy, Office of Science, HEP User Facility. Fermilab is managed by Fermi Research Alliance, LLC (FRA), acting under Contract No. DE-AC02-07CH11359.
These resources included support for the MINERvA construction project, and support
for construction also
was granted by the United States National Science Foundation under
Award No. PHY-0619727 and by the University of Rochester. Support for
participating scientists was provided by NSF and DOE (USA); by CAPES
and CNPq (Brazil); by CoNaCyT (Mexico); by Proyecto Basal FB 0821, CONICYT PIA ACT1413, and Fondecyt 3170845 and 11130133 (Chile); 
by CONCYTEC (Consejo Nacional de Ciencia, Tecnolog\'ia e Innovaci\'on Tecnol\'ogica), DGI-PUCP (Direcci\'on de Gesti\'on de la Investigaci\'on  - Pontificia Universidad Cat\'olica del Peru), and VRI-UNI (Vice-Rectorate for Research of National University of Engineering) (Peru); NCN Opus Grant No. 2016/21/B/ST2/01092 (Poland); by Science and Technology Facilities Council (UK); by EU Horizon 2020 Marie Skłodowska-Curie Action.  \postsubmissionAdd{D.~Ruterbories gratefully acknowledges support from} a Cottrell Postdoctoral Fellowship, \postsubmissionStrike{ from the} Research Corporation for Scientific Advancement \postsubmissionAdd{award number 27467 and National Science Foundation Award CHE2039044}.  We thank the MINOS Collaboration for use of its near detector data. Finally, we thank the staff of
Fermilab for support of the beam line, the detector, and computing infrastructure.

\end{acknowledgments}

\clearpage

\bibliography{MINERvA}

\end{document}